\documentstyle [12pt]{article}
\begin{document}

\hfill {WM-99-108}
\input epsf

\hfill {\today}

 \baselineskip 24pt
{
\Large 
   
  \vskip .2in
   \centerline{The Triple-Alpha Process and the Anthropically}
\centerline{
Allowed Values of the Weak Scale} }
 \vskip .2in
\def\bar{\overline}
 
\centerline{Tesla E. Jeltema and Marc Sher } 
\bigskip
\centerline {\it Nuclear and Particle Theory Group}
\centerline {\it Physics
Department}
\centerline {\it College of William and Mary, Williamsburg, VA
23187, USA}
\vskip .4 in
\centerline{ABSTRACT}

{\narrower\narrower In multiple-universe models, the constants of nature may have
different values in different universes.  Agrawal, Barr, Donoghue and Seckel have
pointed out that the Higgs mass parameter, as the only dimensionful parameter of the
standard model, is of particular interest.  By considering a range of values of this
parameter, they showed that the Higgs vacuum expectation value must have a magnitude less
than $5.0$ times its observed value, in order for complex elements, and thus life, to form.
In this report, we look at the effects of the Higgs mass parameter on the triple-alpha
process in stars.  This process, which is greatly enhanced by a resonance in Carbon-12, is
responsible for virtually all of the carbon production in the universe.  We find that the
Higgs vacuum expectation value must have a magnitude greater than $0.90$ times its observed
value in order for an appreciable amount of carbon to form, thus significantly narrowing
the allowed region of Agrawal et al.}

\newpage

The anthropic principle\cite{anthropic} states that the parameters
of our Universe must have values which allow intelligent life to
exist.  It is a principle which has existed in some form or another
since the beginning of human history.  It has countless
formulations, many of which have religious overtones. 
In recent years, however, the anthropic principle has been revived as a
method of explaining some fine-tuning problems.  For example,
Weinberg has considered\cite{weinberg} whether the principle can
address the relative smallness of the cosmological constant.

In its weak form, the anthropic principle states that, because we
are here to observe them, the observed properties of the universe
must have values which allow life to exist.  This may seem somewhat
obvious or circular, but it becomes significant in some physical
theories which support the existence of domains in the universe in
which different parameters are applicable.  In chaotic inflation
models\cite{chaotic}, for example, different domains may have different
Higgs vacuum expectation values. 
These domains can be regarded as different universes. 
Alternatively,  in  regions of high gravitational
 curvature, new universes may, in some models,  ``pop" out of the
vacuum; these new universes  may have different values of the parameters.  Thus,
considering how our universe (and the life therein) would evolve if the parameters of the
standard model were changed may be physically relevant.

The standard model has (including neutrino masses and mixing) some
24 parameters.   Thus, any complete study of the anthropic
principle would involve study of a complex 24-dimensional parameter
space.  In two recent papers\cite{barrprl,barrprd}, Agrawal, Barr,
Donoghue and Seckel(ABDS) noted that the Higgs mass-squared
parameter is of special interest.    It is the only dimensionful
parameter in the model, and multiple-universe models may be more
likely to have varying dimensionful couplings  than varying dimensionless
ones.  The Higgs mass-squared parameter is also unnaturally small compared with the
parameters of more general theories, such as grand unified theories.  

ABDS considered the range of anthropically allowed values of the
Higgs mass-squared parameter, $\mu^2$.  They considered values of
this parameter ranging from $-M_{Pl}^2$ to $M_{Pl}^2$, where
$M_{Pl}$ is the Planck scale, and we define the sign of $\mu^2$ to
be negative in the standard model.   ABDS considered both the cases $\mu^2<0$ and
$\mu^2>0$.  In the latter case, the electroweak gauge symmetry is still broken by 
quark condensation ($\langle\bar{q}q\rangle\neq 0$).   For the $\mu^2<0$ case, they found
that as one increases the Higgs vacuum expectation value $v\equiv
\langle\phi\rangle=\sqrt{{-\mu^2\over\lambda}}$ from its standard model value, $v_0$, the
first major effect occurs when the deuteron becomes unbound.  This occurs when $v/v_0$
reaches a value of $1.4-2.7$, depending on the nuclear physics model, and is due to the
increasing neutron-proton mass difference.   When $v/v_0$ is greater than about $5.0$, all
nuclei become unstable.  They argue, therefore, that one must have $v/v_0 < 5.0$ (and
possibly less than $2.7$) in order for complex elements to form, and thus life.   They also
note that for
$v/v_0 > 10^3$, the $\Delta^{++}$ becomes stable relative to the proton, leading to a very
unusual universe indeed.   For $\mu^2>0$, the weak scale becomes of the order of magnitude
of the QCD scale, and chemical and stellar evolution become much more complicated.

One process not considered by ABDS is the triple-alpha process in
stars.  This process occurs when two alpha particles first fuse into beryllium
(${}^4He+{}^4He\rightarrow {}^8Be$).  The beryllium has a very short lifetime (of order of
$10^{-16}$ seconds), but lives long enough for further interaction with a third alpha
particle (${}^4He+{}^8Be\rightarrow {}^{12}C^*$) to produce carbon.   Virtually all of the
carbon in the universe is produced through this process.
This process is anthropically significant\cite{anthropic} because it
depends very precisely on the existence of a $0^+$ resonance $7.6$ MeV above the ground
state in ${}^{12}C$.   The existence of this resonance was one of the first, major
successful predictions of astrophysics; being predicted by Hoyle\cite{hoyle} long before
the discovery of the resonance.   Without this resonance, little carbon will be produced. 
Without carbon, it is difficult to see how life could spontaneously develop.   Life, as we
generally define it, requires the existence of a molecule capable of storing large amounts
of information, and it is impossible for hydrogen and helium to form such molecules.
Since the existence of the resonance is a very sensitive function of the parameters of the
model\cite{anthropic}, one might expect it to give much more stringent bounds on $v/v_0$
than those obtained by ABDS.   In this Brief
Report, we examine the dependence of this process on $\mu^2$, and
significantly narrow the range found by ABDS.\   

There have been several calculations concerning the anthropic significance of the
triple-alpha process.  Livio, et al.\cite{livio} calculated the sensitivity of the amount of
carbon production to changes in the location of the $0^+$ resonance, but did not address
the underlying physics behind the location of the resonance.  Oberhummer, et
al.\cite{oberhummer} then did a detailed nuclear physics calculation of the sensitivity of
the location of the resonance to the strength of the nucleon-nucleon potential.  This
required considering several different models for the nuclear reaction rates.  They found
that a change of only a part in a thousand in the strength of the nucleon-nucleon
interaction will change the reaction rate of the triple-alpha process by roughly a factor
of $20$, and a change of two parts in a thousand changes it by roughly a factor of  $400$.

The strength of the nucleon-nucleon interaction, however, is a very complicated function of
the many parameters of the standard model.   Our objective is to relate this strength to
changes in the vacuum expectation value of the Higgs boson, $v$.  Changing $v$ will change
the quark masses, and will also change the value of the QCD
scale.  Both of these are addressed by ABDS.  The quark masses change in a very predictable
way:  $m_q \sim (v/v_0)$.    The QCD scale, $\Lambda$, which is sensitive to the quark
masses through threshold effects (it is assumed that the high energy value is unchanged),
is found by ABDS to scale as $(v/v_0)^\zeta$, where $\zeta$ varies between $0.25$ and
$0.3$---we will take it to be $0.25$ in this work.   From these variations, one can
calculate the variation of the relevant baryon and meson masses, and convert that into an
effect on the strength of the nucleon-nucleon interaction.

The phrase ``strength of the nucleon-nucleon interaction" is, of course, somewhat
ambiguous.  Oberhummer, et al.\cite{oberhummer}, simply multiplied the interaction by a
constant.  When the meson and baryon masses change, however, the entire shape of the
potential changes.   A precise analysis would necessitate using this full potential in the
calculation of the triple-alpha process.  However, these calculations use
``phenomenological" parameters, which are experimentally determined, and the variation
of these parameters with $v$ is unknown.  We therefore
estimate the size of the effect by finding an ``average"
value of the potential, defined as
\begin{equation}
\langle V\rangle={\int_0^\infty\ V(r)|\psi(r)|^2\ d^3r \over
\int_0^\infty\ |\psi(r)|^2\ d^3r}
\end{equation}
where $\psi$ is the two-nucleon wavefunction, obtained by solving the Schr\"{o}dinger
equation, and compare this with Oberhummer, et al.\begin{figure} [h]
\centerline{ \epsfysize 3in \epsfbox{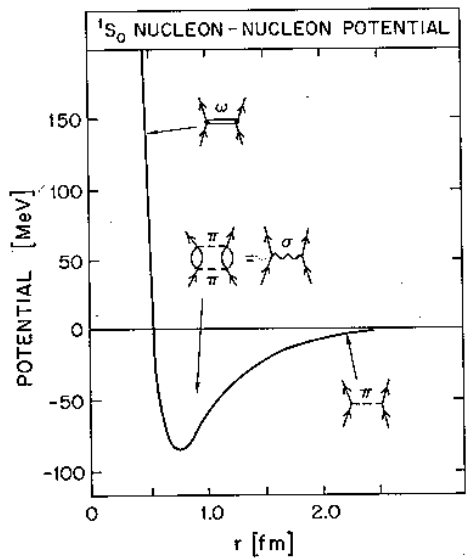}  }
\caption{The nucleon-nucleon potential, from Ref. \cite{ulf}.  The $\sigma$ is believed to
be a two-pion resonance, although it may be a real, but very broad, physical state.}
\end{figure}
 We now have to determine the dependence of the potential on $v/v_0$.

The nucleon-nucleon potential has three main features, shown in Figure 1. 
 There is a
repulsive core, an attractive minimum and a long-range tail from one-pion exchange.  We
will look at two different models for the nucleon-nucleon potential.  The first considers
the repulsive core to be due to the exchange of the $\omega$ vector meson, and the
attractive minimum to be due to the exchange of the hypothetical sigma meson.  Controversy
exists as to whether the sigma meson is an actual particle with a large width, or simply a
correlated two-pion exchange.  We will assume the latter for the moment, but will show that
the results will not change significantly in either case.   The potential can then be
written as
\begin{equation}
V(r)={g_\omega \exp^{-m_\omega r}\over r}-{g_\sigma \exp^{-m_\sigma r}\over r}-{g_\pi
\exp^{-m_\pi r}\over r}
\end{equation}
where the $g_i$, arising from the strong interaction van der Waals forces, are assumed to
be independent of the weak scale.  To find the dependence of $V(r)$ on $v/v_0$, we now need
to ascertain the dependence of $m_\omega$, $m_\sigma$ and $m_\pi$ on $v/v_0$ (as well as
the dependence of the nucleon mass, due to the input into the Schr\"{o}dinger equation).

The dependence of the pion mass on the weak scale is easily determined from the formula
from chiral symmetry breaking, which gives $m_\pi^2\propto f_\pi(m_u+m_d)$.  Since $f_\pi$
varies as $\Lambda_{QCD}$, which varies as $(v/v_0)^\zeta$, and $m_u+m_d$ varies as
$v/v_0$, one can see that $m_\pi \sim (v/v_0)^{1+\zeta\over 2}$.   The nucleon and the
$\omega$ primarily get their masses from QCD, which scale as $\Lambda_{QCD}$, but have
small contributions from the current quark masses.  In MeV, the masses are given by
$m_{nucleon}= 921(v/v_0)^\zeta+18(v/v_0)$ and $m_\omega=768(v/v_0)^\zeta+14(v/v_0)$, where
we have taken the up and down current quark masses to be $4$ and $7$ MeV, respectively.

The mass of the sigma is a different matter, since it is a two-pion correlated state.  We
follow the work of Lin and Serot\cite{serot}, who derive the mass of the $\sigma$ in terms
of the pion mass, the nucleon mass and the pion-nucleon coupling constant.  By varying the
masses of the pion and nucleon in their expressions, we find that $m_\sigma \sim
(v/v_0)^{0.26}$.  This is not a surprising result.  The $\sigma$ mass turns out to be very
insensitive to the pion mass, and thus it can only scale as the nucleon mass, which scales
as $(v/v_0)^{0.25}$.  It also indicates that the result is not significantly changed if one
regards the $\sigma$ to be a real particle, since one would expect such a particle to scale
as the QCD scale, and $\Lambda_{QCD}\sim (v/v_0)^{0.25}$.

With the mass dependences, we now determine the strength of the nucleon-nucleon potential
as $v$ is varied.   It is found that a $1\%$ change in $v$ affects the strength of the
potential by $0.4\%$ (in the same direction); a $10\%$ change in $v$ affects it by $4\%$.
To see how robust this result is, we also considered a completely different nucleon-nucleon
potential, due to Maltman and Isgur\cite{isgur}, using six-quark states.  There are
two parts to the potential, a modified one-pion exchange part and a part due to residual
quark-quark interactions.  The latter, which is most relevant for this analysis, is
entirely due to QCD, and thus its variation with $v$ only depends on the variation through
$\Lambda_{QCD}$, which is determined dimensionally.  The result is similar; a $1\%$
decrease in $v$ decreases the strength of the potential by $0.6\%$.

Now that we have related the strength of the nucleon-nucleon potential to the dependence on
$v$, we can go to the work of Oberhummer et al. who relate that to the rate of carbon
production.  Oberhummer et al. found that a decrease of $2-4\%$ in the strength of the
nucleon-nucleon potential leads to the virtual elimination of carbon production (Livio, et
al.\cite{livio} analyzed both 5 and 20 solar mass stars, although the result is insensitive
to the precise stellar mass).   Comparing with our result from the previous paragraph, we
find that (conservatively taking a $4\%$ decrease as our limit as well as the first
potential model) one must have
$v/v_0$ greater than
$0.90$.   This substantially narrows the region found by ABDS, which had no effective lower
bound on $v/v_0$, but only an upper bound of between $1.4$ and $5$.

How accurate is this result?  As noted earlier, a precise determination of the effects of
changing $v$ on the rate of carbon production in stars would require solving the
twelve-body problem with a varying nucleon-nucleon potential (not to mention three-body
forces).  Oberhummer et al. just varied the overall strength of the two-body potential.  A
full analysis does not seem possible at this time.  We have related the change in the
potential caused by the variation of $v$ to an ``average" potential strength.  This
``mean-field" approach is not particularly precise, but is probably the best that can be
done at this time, given our lack of understanding of nuclear dynamics. The fact that two
very different models of the potential give a similar bound is encouraging.  Thus, our bound
should be taken as a reasonable approximation to the bound that could be obtained with a
full understanding of the nuclear physics involved.

We thank Dirk Walecka and Nathan Isgur for many useful conversations about the
nucleon-nucleon potential, and Eric Dawnkaski for help with the computational aspects of
this work.  This work was supported by the National Science Foundation PHY-9900657.

 \def\prd#1#2#3{{\rm Phys. ~Rev. D}{\bf #1}, #3 (19#2)}
\def\plb#1#2#3{{\rm Phys. ~Lett. B}{\bf #1}, #3 (19#2) }
\def\npb#1#2#3{{\rm Nucl. ~Phys. A}{\bf #1}, #3 (19#2) }
\def\prl#1#2#3{{\rm Phys. ~Rev. ~Lett. ~}{\bf #1}, #3 (19#2) }

\bibliographystyle{unsrt}

\end{document}